\documentclass[3p]{elsarticle}
\usepackage{hyperref}
\usepackage[mathlines,pagewise]{lineno}
\modulolinenumbers[5]

\usepackage{graphicx}
\usepackage{multicol}
\usepackage{bm}
\usepackage{color}

\journal{    }

\begin{document}

\begin{frontmatter}

\title{Diffusion of innovations in finite networks: effects of heterogeneity, clustering, and bilingual option on the threshold in the contagion game model}

\author[JC]{Jeong-Ok Choi}

\author[UY]{Unjong Yu\corref{mycorrespondingauthor}}
\cortext[mycorrespondingauthor]{Corresponding author}
\ead{uyu@gist.ac.kr}

\address[JC]{Division of Liberal Arts and Sciences, Gwangju Institute of Science and Technology, Gwangju 61005, South Korea}
\address[UY]{Department of Physics and Photon Science, Gwangju Institute of Science and Technology, Gwangju 61005, South Korea}

\date{\today}

\begin{abstract}
The contagion threshold for diffusion of innovations is defined and calculated
in finite graphs (two-dimensional regular lattices, regular random networks (RRNs), and two kinds of scale-free networks (SFNs))
with and without the bilingual option.
Without the bilingual option, degree inhomogeneity and clustering enhance the contagion threshold in non-regular networks
except for those with an unrealistically small average degree. 
It is explained by the friendship paradox and detour effect.
We found the general boundary of the cost that makes the bilingual option effective.
With a low-cost bilingual option, among regular lattices, SFNs, and RRNs,
the contagion threshold is largest in regular lattices
and smallest in RRNs.
The contagion threshold of regular random networks is almost the same as that 
of the regular trees, which is the minimum among regular networks.
We show that the contagion threshold increases by clustering
with a low-cost bilingual option.
\end{abstract}


\end{frontmatter}


\section{Introduction}

\textit{Diffusion of innovations} describes the diffusion process of a new idea, technology, or product through
a social network \citep{Rogers03,Kleinberg07}.
The rapid developments of communication,  
transportation technologies, and social networking services
are making connections between people denser and more complex \citep{Lazer09},
and so the mechanism of diffusion of innovations
is becoming more and more complex.

In the diffusion of innovations,
people tend to adopt the innovation when the exposure, which is the proportion of innovation adopters in their neighborhood,
is larger than a certain threshold \citep{Valent96,Centola10,Pei13}.
It is because people usually judge the benefit of adopting new innovation based on the exposure.
As a special case of diffusion of innovations, the epidemic begins from one or a few instigators (or infected people) and
diffuses through a network making all the agents infected when the process is finished. In this paper, our main interest is the epidemics. 

This problem was modeled mathematically as a contagion game by Morris \citep{Morris}.
In this contagion game model, every agent takes one option from two strategies: $A$ and $B$. During the game an agent can get (positive) payoff only when it has the same option as its neighbor; they get the payoff $(1-q)$ and $q$ when they have the strategy $A$ and $B$, respectively. 
To begin with, Morris considered an infinite graph where all the agents have strategy $B$ (status quo option) except for only finite number of instigators that adopt strategy $A$ (innovative option) and keep their strategies $A$.
The innovative option $A$ is assumed to be better than the status quo option ($q<0.5$)
because there is no incentive to transform to the innovative option otherwise.
All the agents but instigators change their strategies to maximize their payoffs
in their local environments (best-response update) infinite times.
It is called that the strategy $A$ become \textit{epidemic} on the graph in the game when the whole graph adopts the innovative option in the end by this dynamics.
The contagion threshold is defined to be the largest
$q$ value that can make the game on the graph epidemic from all the possible finite-size instigators \citep{Morris}.

Another important aspect of a contagion game
is the possibility of the bilingual option, which is compatible with both options
by paying an extra cost. 
Immorlica {\it et al.} extended the contagion game of Morris
by including the bilingual option,
which is represented by the payoff matrix of
Table~\ref{Table_payoff} \citep{Immorlica07}.
This bilingual option makes the dynamics in a contagion game more complex. 
Since it costs $c$ for an agent to choose the bilingual option (i.e., the cost is $c/\Delta \times \Delta = c$ if it has $\Delta$ neighbors), choosing the bilingual option is likely to give the agent larger total payoffs than choosing $A$ or $B$ when $c$ is low enough.
A bilingual option with low cost may help a smooth transition to the innovative option $A$ for an agent with more neighbors. As a result, the contagion threshold increases.
\cite{Immorlica07} obtained contagion threshold $q_0^c$ as a function of cost $c$ in some infinite regular graphs. 
Recently, it was proved that the infinite regular tree has the smallest contagion threshold
among infinite regular graphs of the same degree for every positive $c$ \citep{Yu_Choi17}.

In this paper, we extend the study of \cite{Morris,Immorlica07,Yu_Choi17}
to finite and irregular graphs.
We define the contagion threshold for finite graphs and calculate it
for regular lattices, regular random networks (RRNs), and two kinds of scale-free networks (SFNs) 
with and without the bilingual option.
To see the effect of each network parameter more clearly, we restrict this work to model networks.
Effects of degree heterogeneity, clustering, and bilingual option on the contagion threshold are discussed.

\begin{table*}[tb]
\caption{\label{Table_payoff} Payoff matrix of the agent
    in the contagion game with the bilingual option (option $AB$).
    $q$ is assumed to be smaller than half.
    $c$ is the cost for the bilingual option and $\Delta$ is the degree of the agent.}
\begin{tabular}{ccccc}
\hline
 & & \multicolumn{3}{c}{Strategy of a neighbor of the agent} \\ \cline{3-5}
 & & option $A$ & option $B$ & option $AB$ \\
\hline
                      & option $A$  & $1-q$          & $0$          & $1-q$ \\
Strategy of the agent & option $B$  & $0$            & $q$          & $q$ \\
                      & option $AB$ & $1-q-c/\Delta$ & $q-c/\Delta$ & $1-q-c/\Delta$ \\
\hline
\end{tabular}
\end{table*}

\section{Model and methods}
In a contagion game on an infinite graph with the payoff matrix in Table~\ref{Table_payoff}, the {\it contagion threshold} is defined the maximum value of $q$ to be epidemic. Initially in the game, some agents are instigators and take $A$ while all the rest of agents take $B$. Note that with a bilingual option, the contagion threshold is a function of $c$. 
In the definition of the contagion threshold for infinite graphs,
there is no restriction in the set of instigators as long as its size is finite \citep{Morris,Immorlica07,Yu_Choi17}.
If the same definition is used for finite graphs,
then the contagion threshold will be always 1 with a trivial set of instigators--the entire vertex set.
Another problem is that it is practically impossible to find
the optimal set of instigators that maximize diffusion of innovations in complex networks \citep{Kempe03}.
Therefore, we define the contagion threshold for finite irregular graphs
as follows:
in a {\it finite} graph, the contagion threshold $q_0$ is the expectation
value of maximum $q$ for the innovative option $A$ to spread the whole graph
from a state in which all nodes adopt the option $B$
except {\it one node} (an instigator) that adopts and keeps the option $A$,
through the contagion game that all nodes except the instigator
perform best-response update as many times as possible.
This definition is justified when the density of instigators is low
and each instigator acts independently of the other instigators.
We denote $q_0^c$ the contagion threshold for finite graphs with the bilingual option of cost $c$.
This definition gives the same value as that by the original definition for the infinite lattice in the square lattice, but they may give different results in other networks.

\begin{figure}[tb]
\includegraphics[width=10.0cm]{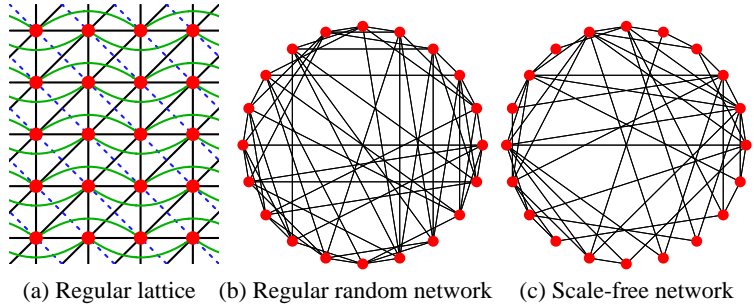}
\caption{Regular lattice network, regular random network, and Barab{\'a}si-Albert scale-free network.
Red circles and lines represent nodes and edges, respectively. 
In (a), black solid lines represent edges of lattice of $\Delta \geq 6$;
blue dashed lines, $\Delta \geq 8$; and  green curves, $\Delta = 10$.
$\Delta$ is the degree of each agent.}
\label{Fig_lattice}
\end{figure}

Three kinds of two-dimensional regular lattices with a degree ($\Delta$) of 6, 8, and 10 
are studied in this work. (See Fig.~\ref{Fig_lattice}(a).)
The lattices of $\Delta=6$ and $\Delta=8$ are the same as
the triangular lattice and the square lattice with Moore neighborhood, respectively.
Two third-nearest neighbors are added to the lattice of $\Delta=8$
to make a lattice with $\Delta=10$.
The periodic boundary condition is used. The {\it clustering coefficient} ($\xi$) of a graph is the ratio of the number of triangles over the number of connected triples of nodes. 
The clustering coefficients are $\xi=2/5$, $3/7$, and $7/15$
for $\Delta=6$, $8$, and $10$, respectively.

RRNs with $N$ nodes of degree $\Delta$ 
were generated by the Steger-Wormald algorithm \citep{Steger99}. 
It is faster than the original pairing model \citep{Bollobas80} 
and generates uniform RRNs asymptotically
when $N$ is large \citep{Kim03}.
The algorithm begins by making a set of unpaired arms of nodes
$\{ (1,1),$ $(1,2), \cdots, (1,\Delta),$ $(2,1), \cdots, (N,\Delta-1),$ $(N,\Delta)\}$,
whose element $(a,b)$ represents the $b$-th unpaired arm of node $a$.
Two elements $(i_1,i_2)$ and $(j_1,j_2)$ are chosen at random from the set.
If $i_1 \neq j_1$ and the two nodes have not been
connected before, then nodes $i_1$ and $j_1$ are connected.
After connecting $i_1$ and $j_1$, the two elements $(i_1,i_2)$ and $(j_1,j_2)$ are eliminated from the set of unpaired arms.
This pairing procedure is repeated until no unpaired arm is left.
The probability that this algorithm fails
to make a $\Delta$-regular network without multiple edges is very small.
The clustering coefficient of RRNs is $(\Delta-1)/N$ asymptotically, which approaches zero for large $N$ (see the appendix). 

SFNs were constructed using the growing method \citep{Barabasi99}.
Starting from a network of $m$ isolated nodes,
a new node is introduced and $m$ edges connect the new node and existing nodes avoiding multiple edges. 
This growth process (i.e., an addition of one node and $m$ edges to the network) 
is repeated until the network has $N$ nodes. 
The node to connect to a new node is chosen by two algorithms:
preferential attachment (PA) and triad formation (TF) \citep{Holme02}.
As for the PA, the probability to be selected is proportional to its degree,
while the node is chosen randomly among
neighbors of the node chosen immediately before by the PA, in the TF.
For the first edge of a new node, the PA is used,
and from the second connection, TF is chosen with probability $P_{\mathrm{TF}}$ if it is possible.
The Barab{\'a}si-Albert SFN, whose clustering coefficient vanishes for large network size \citep{bollobas03}, is obtained with $P_{\mathrm{TF}}=0$.
The clustering coefficient can be enhanced by adjusting $P_{\mathrm{TF}}>0$; 
it is called the Holme-Kim SFN.
The two SFNs have almost the same degree distribution, which is $P(\Delta) \sim \Delta^{-3}$ \citep{Holme02}.
The average degree of each SFN is $\langle \Delta \rangle = 2m$.

As for regular lattices, where network structure is unique and all nodes are symmetric,
the contagion threshold $q_0^c$ can be calculated analytically.
For other networks, it was obtained numerically.
The location of the instigator can be crucial, especially, in SFNs.
We confirmed that high-degree instigators tend to give higher $q_0^c$, but
there are exceptions; it is known to be NP-hard to determine an optimal set of instigators that makes the epidemic process most effective \citep{Kempe03}.
The contagion threshold for each instigator was calculated and averaged
to determine $q_0^c$ of the network.
Ten kinds of networks were generated for each case, and the results of them were averaged.
Differences in the results by different implementations of networks are very small.

\section{Results and discussion} 

\begin{figure*}[tb]
\includegraphics[width=13.0cm]{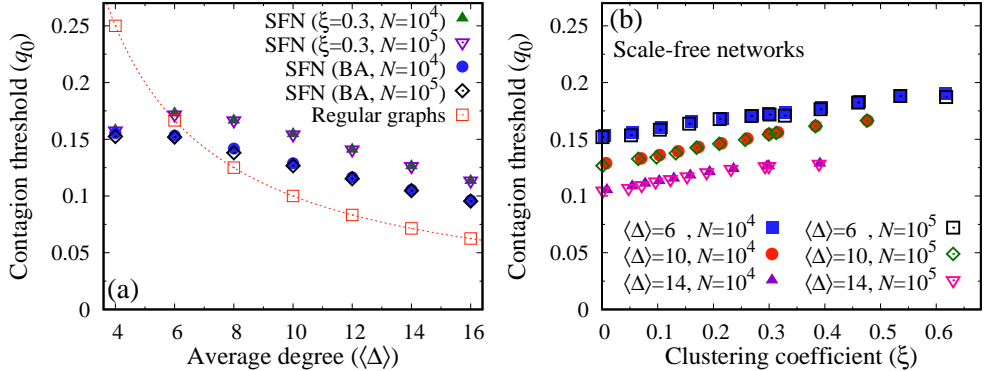}
\caption{Contagion threshold $q_0$ without the bilingual option for regular graphs and scale-free networks (SFNs).
The clustering coefficient ($\xi$) of the SFN
is controlled by $P_\mathrm{TF}$ in the Holme-Kim's algorithm while that of the Barab\'{a}si-Albert (BA) SFN
is almost zero. A red dotted curve in (a) is $1/\Delta$.
Error-bars are similar as the symbol size except those of regular graphs, which are exact.}
\label{Fig_AB}
\end{figure*}

\subsection{Contagion threshold without bilingual option}

The contagion threshold $q_0$ without bilingual option ($c=\infty$) is shown in Fig.~\ref{Fig_AB}.
Size dependence on $q_0$ is very small if it exists.
We consider the effects of degree heterogeneity and clustering, which is not conclusive yet in spite of
various studies \citep{Watts2002,Montanari10,Young11,Acemoglu11,Coupechoux2011,Coupechoux2014}.

A node $v_i$ having only one neighbor that had adopted innovative option $A$
changes its strategy into $A$ if and only if $q<1/d(v_i)$, where $d(v_i)$ is its degree.
Therefore, in regular graphs such as regular lattices and RRNs,
the contagion threshold $q_0$ is $1/\Delta$.
In SFNs, $q_0$ also decreases with average degree $\langle\Delta\rangle$,
but the decrement is smaller than regular graphs,
and $q_0$ is higher than $1/\langle\Delta\rangle$ for $\langle\Delta\rangle \geq 8$.
We confirmed that SFNs have higher $q_0$ than regular graphs at least up to $\langle\Delta\rangle=50$.
Thus, the contagion threshold is positively correlated with
the degree heterogeneity in the two families of graphs except for networks of
extremely small average degree ($\langle\Delta\rangle < 6$).
According to \citep{Dunbar93}, humans have about 150 relations on average at any given time, among which about 12-20 relations are very intimate (sympathy group) \citep{Zhou05}.
Therefore, we conclude that degree heterogeneity promotes the epidemics in contagion games for realistic cases.
In \cite{Watts2002}, however, it was shown that classical random networks, which have the Poisson degree distribution, have larger contagion threshold than SFNs for average degree $1 < \langle \Delta \rangle \le 30$.
This discrepancy of the results in \citep{Watts2002} and our result is not necessarily contradictory, and we claim that it is mainly due to the structural differences in SFNs.

Heterogeneity of networks affects the epidemic process in two ways. 
The first one is the {\it friendship paradox effect}. 
The instigator is chosen at random and its average degree is $\langle \Delta \rangle$,
but the average degree of its neighbor $\langle \Delta' \rangle$ is larger than $\langle \Delta \rangle$ in non-regular networks: 
\begin{eqnarray}
\langle \Delta' \rangle &=& \frac{\sum_{v \in V(G)} \left[d(v)\right]^2}{\sum_{v \in V(G)} d(v)} 
= \frac{\sum_{v \in V(G)} \left[d(v)\right]^2}{2e} 
= \langle \Delta \rangle + \frac{\sigma^2}{ \langle \Delta \rangle} ,
\end{eqnarray}
where $e$ is the number of links and $\sigma$ is the standard deviation of the degree distribution of the graph \citep{Feld91}.
Therefore, it is harder to persuade its neighbors in a highly-heterogeneous network, 
whose standard deviation of the degree distribution is large.
On the other hand, heterogeneity may promote the contagion process through the mechanism that we call the {\it detour effect}.
On average, it is harder for the instigator to persuade its neighbors by the friendship paradox effect,
but some neighbors of the instigator may have a degree smaller than $\langle \Delta \rangle$
and are easy to persuade; the contagion process can continue through these channels.
Neighbors with a large degree are hard to persuade,
but it is possible to persuade them later with the help of other paths from the instigator to the nodes.
Therefore, heterogeneity may enhance the contagion process when there are cycles.
The detour effect becomes dominant only for high degree nodes, and so
the enhancement of the contagion threshold by heterogeneity is effective 
only in networks with a high average degree as shown in Fig.~\ref{Fig_AB}(a).
The most probable degree is $\langle \Delta \rangle/2$ in the SFNs 
made by the growing method in this work.
To the contrary, the SFNs used in \citep{Watts2002} are made by the configuration model 
and the most frequent value of the degree is $\Delta=1$; 
therefore, the detour effect is ineffective
and heterogeneity suppresses the contagion threshold through the friendship paradox effect.



Figure~\ref{Fig_AB} shows a positive correlation between the clustering coefficient and contagion threshold on SFNs:
when the average degree is fixed, $q_0$ is higher in Holme-Kim SFNs than in Barab\'{a}si-Albert SFNs.
Interestingly, this effect vanishes in networks with an extremely small average degree, 
consistently with \citep{Coupechoux2014}. 
It can be understood also by the detour effect in heterogeneous networks: 
the detour effect is reinforced by short cycles, and so clustering increases the contagion threshold
in heterogeneous networks
except when the average degree is extremely small.


\subsection{Contagion threshold with bilingual option}

\begin{figure*}[tb]
\includegraphics[width=13.0cm]{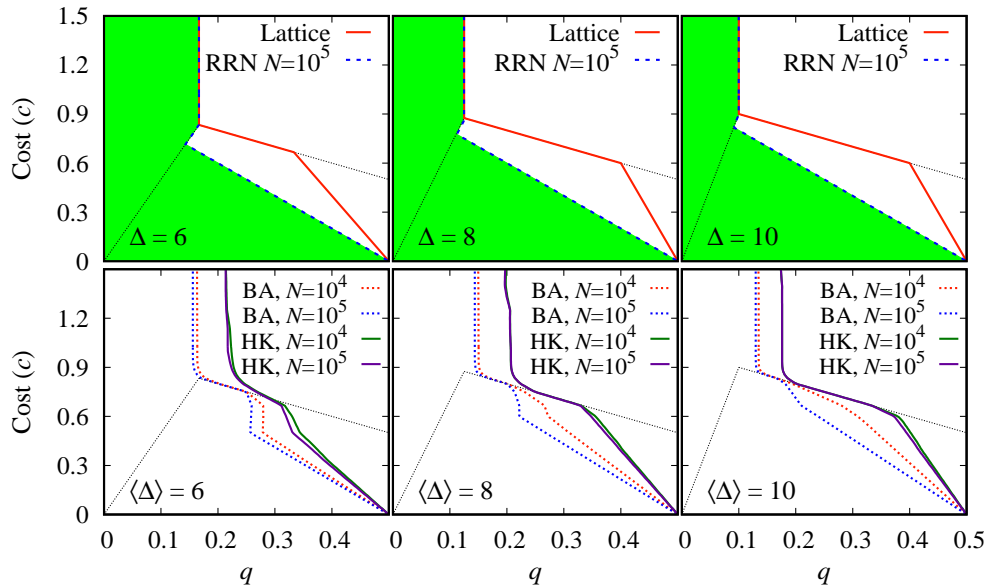}
\caption{Contagion threshold with the bilingual option of cost $c$.
Upper panels are for regular graphs: regular lattices and regular random networks (RRN).
The green polygons represent the Contagion threshold of the $\Delta$-regular tree.
Lower panels are for scale-free networks (SFNs) in the Barab\'{a}si-Albert (BA) algorithm ($\xi=0.0$)
and Holme-Kim (HK) algorithm ($\xi=0.3$).
Black dotted straight lines separate high-cost region 
($c>(\langle\Delta\rangle - 1)q$ or $c > 1-q$)
and low-cost region ($c < (\langle\Delta\rangle - 1)q$ and $c < 1-q$).}
\label{Fig_ABAB}
\end{figure*}

In the model with the bilingual option $AB$, Fig.~\ref{Fig_ABAB} presents the contagion threshold $q_0^c$ for lattices, RRNs, and SFNs. 
In spite of the cost $c$, the bilingual option ($AB$) can be more advantageous since it is compatible with both options $A$ and $B$ resulting in higher payoffs. 
Thus, the contagion threshold of a contagion game with the bilingual option is a function of the cost $c$.

\begin{figure}[tb]
\includegraphics[width=10.0cm]{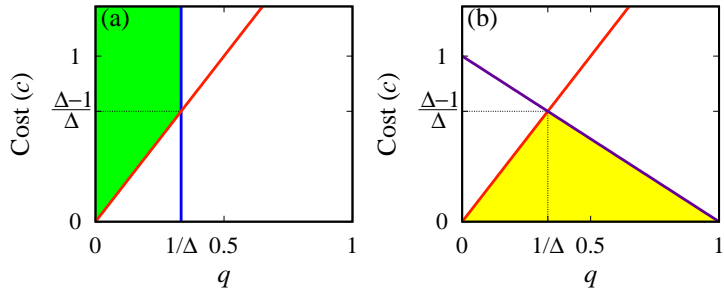}
\caption{
Regions $(q,c)$ for the neighbor of the instigator 
to change its strategy into $A$ in (a) and into $AB$ in (b)
for $\Delta$-regular graphs.
The red, blue, and purple straight lines 
represent $c=(\Delta-1)q$, $q=1/\Delta$, and $c=1-q$.
Yellow and white regions in (b) show the low-cost and high-cost regions, respectively.}
\label{Fig_ABAB_schematic}
\end{figure}

In the $\Delta$-regular finite graph, to be epidemic, at least one of the neighbors of the instigator must choose either $A$ or $AB$ at the first step.
If $q < {1}/{\Delta}$ and $c > (\Delta - 1)q$, then it chooses $A$ (see Fig.~\ref{Fig_ABAB_schematic}(a)). 
In this case, the contagion game continues to be epidemic without the bilingual option.
If $c < (\Delta - 1)q$ and $c < 1-q$, then a neighbor of the instigator chooses $AB$ (see Fig.~\ref{Fig_ABAB_schematic}(b)), 
and the final state can be epidemic or a mixture of the three strategies.
Therefore, the contagion threshold $q_0^c$ is at most $1/\Delta$ 
if $c > ({\Delta-1})/{\Delta}$, and $q_0^c < 1-c$ 
if $c < ({\Delta-1})/{\Delta}$ for a $\Delta$-regular finite graph. 
Interestingly, for various $c$ and $q$ on $\Delta$-regular graphs, there appears qualitative change across lines $c = (\Delta - 1)q$ and $c=1-q$; 
$AB$ becomes effective only if $c$ is low below these lines.
We call this region ($c < (\Delta - 1)q$ and $c < 1-q$) as the {\it low-cost region}. 



Figure~\ref{Fig_ABAB} shows that qualitative change occurs across the boundary of the low-cost region also in non-regular networks as well as in regular graphs. 
Therefore, the boundary is supposed to be general, though it was derived in regular graphs.
In the high-cost region ($c>(\langle\Delta\rangle - 1)q$ or $c > 1-q$), $q_0^c$ is the same as $q_0$ without bilingual option or approaches $q_0$ fast. 
Hence, the bilingual option reflects its effect in the low-cost region ($c < (\langle\Delta\rangle - 1)q$ and $c<1-q$).
In the low-cost region, $q_0^c$ increases as the bilingual cost is reduced except RRNs, 
which have somewhat smaller $q_0^c$ than $q_0$ near the boundary of $c=1-q$.
The decrease of $q_0^c$ means that the bilingual option can hinder the contagion process by protecting the status quo option from extinction.
Contagion threshold is largest in lattices, intermediate in SFNs, and smallest in RRNs.
Interestingly, the contagion threshold of RRNs is almost the same as that of the infinite regular tree, which is proved to have the smallest value among all infinite regular graphs \citep{Yu_Choi17}. 
It is easy to see that $q_0^c$ for a finite regular lattice is as small as the contagion threshold defined for an infinite regular graph with a bilingual option as in \citep{Immorlica07,Yu_Choi17}. 
Therefore, it is highly probable that RRN has relatively small contagion threshold among all finite graphs with $N$ nodes. 
It is understandable because RRN has a tree-like structure and the (average) number of cycles with a fixed length is reasonably small \citep{Bollobas}. 
In other words, both networks have small clustering and no short cycles. 
In contrast to the high-cost region, $q_0^c$ increases with the average degree
in all networks considered in this paper.
Clustering always enhances the contagion threshold.
Size-dependence is negligible except for the Barab\'{a}si-Albert SFNs,
whose $q_0^c$ decreases a little bit as the number of nodes in the network.

\section{Summary}

We defined and calculated the contagion threshold in finite graphs: regular lattices, RRNs, and SFNs.
Without the bilingual option, the contagion threshold $q_0$ in regular graphs is $q_0 = 1/\Delta$
and independent of the network structure.
The contagion threshold in SFNs is larger
than that of regular graphs for average degree $\langle\Delta\rangle\geq 8$.
This can be understood by the friendship paradox effect and detour effect in SFNs.
Clustering in SFNs enhances the detour effect to increase the contagion threshold.
The bilingual option makes a qualitative change of contagion threshold
when the bilingual cost $c$ is lower than $(1-q)$ and $(\langle\Delta\rangle-1)q$.
With the bilingual option of low cost, 
$q_0^c$ is largest in lattices, intermediate in SFNs, and smallest in RRNs.
The contagion threshold of RRNs of degree $\Delta$
is almost the same as the $\Delta$-regular tree, which has the smallest $q_0^c$
among $\Delta$-regular networks.
In both cases of regular networks and SFNs, the innovative option
is easier to spread as the clustering coefficient of the network increases.
We found little network size dependence on the contagion threshold except
in Barab\'{a}si-Albert SFNs with a low-cost bilingual option.

\section*{Acknowledgments}
This work was supported by GIST Research Institute (GRI) grant funded by the GIST in 2019.

\section*{References}
\bibliography{diff_inno}

\appendix

\section{Clustering coefficient in regular random networks}

We consider a $\Delta$-regular random network of $N$ vertices,
which is assumed to be large.
Suppose that a vertex $x$ is connected to vertices $y$ and $z$.
The clustering coefficient $\xi$ is the probability that
vertices $y$ and $z$ are connected.
Among $\Delta$ arms of $y$, one edge connects $x$ and $y$,
and $y$ has $(\Delta-1)$ arms to be connected.
And there are $(N-2)$ vertices that can be connected to $y$
excluding $x$ and $y$.
Therefore, The probability that vertices $y$ and $z$ are not connected $(1-\xi)$ can be calculated as
\begin{eqnarray}
&&1-\xi = \left( \frac{N-3}{N-2} \right)
\left( \frac{N-4}{N-3} \right) 
\left( \frac{N-5}{N-4} \right) 
\cdots
\left( \frac{N-(\Delta+1)}{N-\Delta} \right) \\
&&~~~~~~~ = \left( 1 - \frac{1}{N-2} \right)
\left( 1 - \frac{1}{N-3} \right) 
\left( 1 - \frac{1}{N-4} \right)
\cdots
\left( 1 - \frac{1}{N-\Delta} \right)\\
&&~~~~~~~ \approx 1 - \left(\frac{1}{N-2}\right) - \left(\frac{1}{N-3}\right) 
- \left(\frac{1}{N-4}\right)
- \cdots - \left(\frac{1}{N-\Delta}\right)
\end{eqnarray}
in the limit of $N \gg \Delta$.
Therefore, the clustering coefficient or the probability for $y$ and $z$ to be connected to each other is
\begin{eqnarray}
\xi \approx  \left(\frac{1}{N-2}\right) + \left(\frac{1}{N-3}\right) 
+ \left(\frac{1}{N-4}\right) 
+ \cdots + \left(\frac{1}{N-\Delta}\right) \approx \frac{\Delta-1}{N} .
\end{eqnarray}

\end{document}